\documentstyle[useAMS,graphicx,longtable]{mn2e}

\newcommand{\hMpc}{{\ifmmode{h^{-1}{\rm Mpc}}\else{$h^{-1}$Mpc}\fi}}
\newcommand{\hkpc}{{\ifmmode{h^{-1}{\rm kpc}}\else{$h^{-1}$kpc}\fi}}

\def\approxlt{\mathrel{\spose{\lower 3pt\hbox{$\sim$}}
        \raise 2.0pt\hbox{$<$}}}
\def\approxgt{\mathrel{\spose{\lower 3pt\hbox{$\sim$}}
        \raise 2.0pt\hbox{$>$}}}
\def\approxpropto{\mathrel{\spose{\lower 3pt\hbox{$\sim$}}
        \raise 2.0pt\hbox{$\propto$}}}

\title{A model of AW UMa}
\author[B. Paczy\'nski, R. Sienkiewicz and D.M. Szczygie{\l{}}]
{B. Paczy{\'n}ski$^{1}$\thanks{e-mail: bp@astro.princeton.edu},
 R. Sienkiewicz\thanks{e-mail: rs@camk.edu.pl}
and D.M. Szczygie{\l{}}$^{2}$\thanks{e-mail: dszczyg@astrouw.edu.pl}\\
$^{1}$Princeton University Observatory, Peyton Hall, Princeton, NJ 08544, USA\\
$^{2}$Warsaw University Observatory, Al.Ujazdowskie 4, 00-478 Warsaw, Poland}

\date{Accepted --.
      Received -- ;
      in original form --}

\pubyear{2006}

\begin{document}

\maketitle

\label{firstpage}

\begin{abstract}
The contact binary AW UMa has an extreme mass ratio,
with the more massive component (the current primary) close to the main sequence,
while the low mass star at ${\rm q \approx 0.1 }$ (the current secondary) has a 
much larger radius than a main sequence star of a comparable
mass. We propose that the current secondary has almost exhausted hydrogen
in its center and is much more advanced in its evolution, as
suggested by St\c epie{\'n}. Presumably the current secondary
lost most of its mass during its evolution with part of it transferred to
the current primary. After losing a large fraction
of its angular momentum, the binary may evolve into a system of FK Com
type.
\end{abstract}


\section{Introduction}  
\label{sect:intro1}

Since its discovery (Paczy\'nski 1964), AW UMa has played a crucial role 
in our understanding of contact binaries by forcing all theories to 
explain both, its extremely low mass ratio of ${\rm q \sim 0.08}$, and the 
apparent equality of effective temperatures of both components.
Eclipsing binaries of W UMa type are in direct contact, sharing a
common envelope around both components.  Lucy (1968a,b) was the
first to recognize that the convection of gas within a common envelope 
equalizes the entropy, and hence the effective temperature is 
approximately uniform and the colours of both components remain
constant throughout the orbital phase.  Yet, as the
components have very different masses, most nuclear energy is generated
in the more massive component (the current primary),
 and it has to be redistributed throughout
 the common envelope.  It was shown that this process is unstable
on thermal (Kelvin-Helmholtz) time scale, which leads to
relaxation oscillations, with the matter being transferred from the
more massive component to the less massive star,
 and vice versa (Lucy 1976, Flannery 1976,
Robertson and Eggleton 1977, Yakut and Eggleton 2005, and references 
therein).
If we assume, that the mass ratio reversal took place 
during previous evolution of such systems, the current secondary was 
the initial primary while the current primary was the initial secondary.
Hereafter, we shall use ``the primary'' instead of ``the current primary'' 
and ``the secondary'' instead of ``the current secondary''.

Hazlehurst (1970) was the first to suggest that contact binaries may
be evolved off the zero age main sequence.  St\c epie\'n (2003, 2006)
suggested that the secondary is the
more advanced in its evolution, in analogy with the ``Algol paradox''.
It is not clear if this assertion is generally
correct. We propose this is likely applicable to systems with 
extreme mass ratios,
like AW UMa (Mochnacki and Doughty 1972), because the secondary's
radius is so much larger than it would be on the main sequence. 
This ``radius excess'' is the primary reason to suspect that the
secondary is very advanced in its evolution. 

\section{Parameters of AW UMa}
\label{sect:parameters}

The following parameters of AW UMa were found on the Hipparcos website
in Strasbourg: ${\rm V = 6.90 }$ mag, trigonometric parallax
${\rm \pi = 15.13 }$ mas, spectral type F1. Assuming the bolometric
correction ${\rm BC = -0.1 }$, these are combined to obtain the
absolute magnitude of AW UMa:
$$
{\rm L = 6.61 ~ L_{\odot} }
$$ 
Assuming all the luminosity is due to the primary, our model on the Main 
Sequence gives
$$
{\rm M_1 = 1.61 ~ M_{\odot} , \hskip0.5cm X = 0.7, \hskip0.5cm Z = 0.02,  \hskip0.5cm (by ~ mass)}
$$ 
where ${\rm M_1}$ denotes the mass of the primary component. 
This is consistent with a rather wide range of observational
determinations of the total mass of the system
(1.3 or 1.7 ${\rm M_{\odot}}$ (Rensing et al. 1985), 1.4 or 1.9 ${\rm M_{\odot}}$ 
(Ruci\'nski 1992) which strongly depended on the assumed value of
the mass ratio ${\rm q }$ in the interpretation of the radial velocity data.
The binary period is
$$
{\rm P_{bin} = 0.4387 ~ d \hskip 0.5cm }
$$

For many years the mass ratio was adopted following Mochnacki and Doughty
(1972, Fig. 1) as ${\rm q = 0.08 }$. However, 
new high-quality data obtained at the David Dunlap Observatory 
in 2006 (Ruci\'nski, private communication) suggest that this value
is too low and may be larger, 
${\rm q = 0.1 \pm 0.02}$\footnote{Dr. Ruci\'nski writes that the DDO data
indicate velocity field deviations from the contact model and
will require special investigation.}.
As the new result has not been published yet, we consider three values:
${\rm q = 0.08, 0.10, 0.12 }$. This choice of mass ratio slightly affects the 
size of the low mass secondary's Roche lobe, following Eggleton (1983).

\section{Outline of the problem}
\label{sect:outline}

We propose a model in which the secondary was a star of
${\rm \sim 1.5 ~ M_{\odot}}$.  It evolved off the main sequence, and it was
stripped of most of its mass, down to the present ${\rm 0.14 - 0.18 ~ M_{\odot}}$.
Some mass was transfered to the primary, some was
lost from the binary.  Also, some angular momentum was lost from the binary.  
We assume that a complicated evolution of AW UMa can be approximated with
a model in which the primary has a structure of a single
star, somewhat evolved off the Zero Age Main Sequence (ZAMS), while
the secondary has a structure of a single star evolved up
to the formation of its helium core, and stripped of most of its mass. 
In other words, we approximate the evolution of the two components
of AW UMa with the evolution of two single stars.  

\section{Evolutionary calculations}
\label{sect:evol}

We adopt the initial chemical composition of AW UMa as
${\rm X = 0.7, ~~ Z = 0.02 }$, (by mass), and we use the evolutionary
code as described in the readme file in:

\centerline{ http://ftp.camk.edu.pl/camk/rs/04/readme.04 }

\noindent
This code follows evolution of a single, spherically symmetric star
of a constant mass, in a standard manner. When necessary, the code
was modified to take into account a rapid mass loss (Stage II, see below).
As concerns input physics, we use the Livermore opacities
(OPAL, Iglesias and Rogers, 1996) supplemented with molecular and grain
opacities as given by Alexander and Ferguson (1994).
We use, as well, the Livermore equation of state (Rogers et al., 1996,
Rogers, 2001). Nuclear reaction rates are calculated according to
Bahcall and Pinsonneault (1995) updated according to Adelberger et al.(1998).
This code has been already used in some previous works (e.g., Dziembowski
 et al., 2001).
A grid of stellar models was calculated, with masses 
${\rm M_{2,0}/M_{\odot} = 1.00, ~ 1.28, ~ 1.79 }$, evolved from ZAMS
until all hydrogen was burned out in their cores and helium cores were formed. 
This was referred to as Stage I in our evolution.
During this standard evolution, the initial convective cores of the
1.28 and 1.79 ${\rm M_{\odot}}$ models vanish and - during a short 
interval before the helium isothermal cores are fully formed and while in 
the radiative equilibrium - the rest of the hydrogen is burned out.
The initial mass of the secondary ${\rm M_{2,0}}$ is the first parameter of our grid.
We consider these models as possible progenitors of the secondary of AW UMa.
The only constraint we put on the initial mass of the primary ${\rm M_{1,0}}$ is
that it is significantly lower than ${\rm M_{2,0}}$ to ensure that this star will be
only slightly evolved when the secondary's radius reaches its Roche lobe.
But we do not have to assume an extreme initial mass ratio of the components because a
significant amount of initial mass of the secondary may be lost into the interstellar
medium.

A degree of exhaustion of hydrogen in the interiors of these three
models, ${\rm X_{c,0}}$, is the second parameter of our grid. 
We have only considered ${\rm X_{c,0}}$ as low as 0.0241 (by mass) 
or less (see Table 1).
\begin{table}
\caption{Parameters of evolutionary advance in the initial models of the 
tracks shown in Figs.1-3. ${\rm X_{c,0}}$ denotes central hydrogen content by mass.
${\rm M_{He,0}}$ denotes mass of a helium, i.e., mass of a hydrogen exhausted 
core. Values of all masses are in solar units.}
\begin{center}
\begin{tabular}{c c c c c}
Track & ${\rm M_{2,0}}$ & ${\rm M_{2}}$ & ${\rm X_{c,0}}$ & ${\rm M_{He,0}}$ \\
\hline
     a  &    1.79  &    0.18  &    2.71e-4  &           \\
     b  &    1.79  &    0.18  &    7.31e-3  &           \\
     c  &    1.79  &    0.18  &    2.41e-2  &           \\
     d  &    1.28  &    0.18  &             &    0.0268 \\
     e  &    1.28  &    0.18  &             &    0.0145 \\
     f  &    1.28  &    0.18  &             &    0.0015 \\
     g  &    1.00  &    0.18  &             &    0.1059 \\
     h  &    1.00  &    0.18  &             &    0.0598 \\
     i  &    1.00  &    0.18  &             &    0.0423 \\
\hline
     j  &    1.79  &    0.16  &    1.68e-3  &           \\
     k  &    1.79  &    0.16  &    1.44e-2  &           \\
     l  &    1.79  &    0.16  &    2.41e-2  &           \\
     m  &    1.28  &    0.16  &             &    0.0441 \\
     n  &    1.28  &    0.16  &             &    0.0268 \\
     o  &    1.28  &    0.16  &             &    0.0145 \\
     p  &    1.00  &    0.16  &             &    0.0879 \\
     q  &    1.00  &    0.16  &             &    0.0598 \\
     r  &    1.00  &    0.16  &             &    0.0423 \\
\hline
     s  &    1.79  &    0.14  &    2.71e-4  &           \\
     t  &    1.79  &    0.14  &    7.31e-3  &           \\
     u  &    1.79  &    0.14  &    2.41e-2  &           \\
     v  &    1.28  &    0.14  &             &    0.0268 \\
     w  &    1.28  &    0.14  &             &    0.0145 \\
     x  &    1.28  &    0.14  &             &    0.0015 \\
     y  &    1.00  &    0.14  &             &    0.0255 \\
     z  &    1.00  &    0.14  &   $<$ 1e-6  &  $<$ 1e-7 \\
\hline
\end{tabular}
\end{center}
\end{table}
Adopted values of ${\rm X_{c,0}}$ are 
not explicitly shown in Figs. 1-3 to preserve their clarity.
In these, for each line style representing a different ${\rm M_{2,0}}$, the model 
tracks are shifted towards the left as the amount of hydrogen exhausted during 
Stage I increases.

Next, for each ${\rm M_{2,0}}$, we were stripping mass of the three evolved models
assuming thermal equilibrium - the evolution was frozen, i.e. there were
no time dependent terms. This was Stage II, during which the mass was reduced to
${\rm M_{2}/M_{\odot} = 0.18, ~ 0.16, ~ 0.14 }$, consecutively. 
The value of ${\rm M_{2}}$ is the third parameter of our grid of models. This way, 
following Stage II, we had 27 initial models, which were next evolved 
with no additional mass loss - this was Stage III. We looked at the radii of 
the models, expecting that some will be expanded enough to be acceptable
as a model for the secondary of AW UMa.

The 27 stellar model tracks are our guesses for what the secondary's evolution 
might be like. The primary star was just assumed to have the mass of
${\rm M_{1} = 1.61 ~ M_{\odot} }$, and willing to accept matter from the
secondary. The only possible way for it to influence evolution of the secondary 
is through the value of the secondary's Roche lobe radius, ${\rm R_{L2}}$, 
being a function of ${\rm M_{1}}$ and ${\rm M_{2}}$, for a given binary period, 
${\rm P_{bin}}$.

The mass and angular momentum loss from the system was, for simplicity of the 
argument, ignored. 
But we  may interpret the whole process as analogous to the evolution of 
short-period Algols, as there exist among them
objects with extreme mass ratios and very short orbital periods (e.g., R CMa: 
${\rm M_{2} = 0.17 ~ M_{\odot}, M_{1} = 1.07 ~ M_{\odot}, P_{bin} = 1.14 ~ days}$; 
for more examples see Sarna and de Greve 1996, and Ibano\v{g}lu et al. 2006). 
So the evolutionary scheme for AW UMa could be the one for a short-period Algol, 
such that a progenitor is a detached close binary which lost a fraction of mass 
and angular momentum via magnetized wind during its main sequence life so that
its period is short enough at the beginning of Stage II. An additional mass and
angular momentum loss takes place during nonconservative Stage II.
Nonconservative mass exchange is accepted in many models of Algol formation
(Sarna and de Greve 1996, Nelemans et al. 2000, Eggleton and Kiseleva-Eggleton 2002).

In Table 2 we give a few examples of binaries in which the secondary is
just filling its critical Roche lobe when hydrogen in its core is nearly depleted
(see Table 1). Adopting reasonable values of the mass ratio from the range 0.3-0.8
we obtain values of the orbital period in the range ${\rm P_{bin} = 0.89 - 1.37}$ days.
This is just an example what the initial Stage II set of parameters might be.
As one can see, to obtain the present value of the orbital period of AW
UMa we have to assume further mass and angular momentum loss during the mass transfer
process.

\begin{table}
\caption{Exemplary sets of the initial ``Stage II'' parameters of AW UMa. Here 
${\rm q_{II}}$ denotes the mass ratio of the system and ${\rm P_{bin}}$ denotes 
the orbital period of the binary. See text for further explanations.}
\begin{center}
\begin{tabular}{c c c c c}
${\rm M_{2,0} / M_{\odot}}$ & ${\rm R / R_{\odot}}$ & ${\rm L / L_{\odot}}$ & ${\rm q_{II}}$ & ${\rm P_{bin}} ~ [days]$ \\
\hline
1.79 &  2.6  & 15.4 & 0.3 & 1.366 \\
1.79 &  2.6  & 15.4 & 0.5 & 1.271 \\
1.28 &  2.2  & 5.0 & 0.5 & 1.170 \\
1.28 &  2.2  & 5.0 & 0.7 & 1.099 \\
1.00 &  1.8  & 2.0 & 0.8 & 0.894 \\
\hline
\end{tabular}
\end{center}
\end{table}


\begin{figure}
\includegraphics[width=\linewidth]{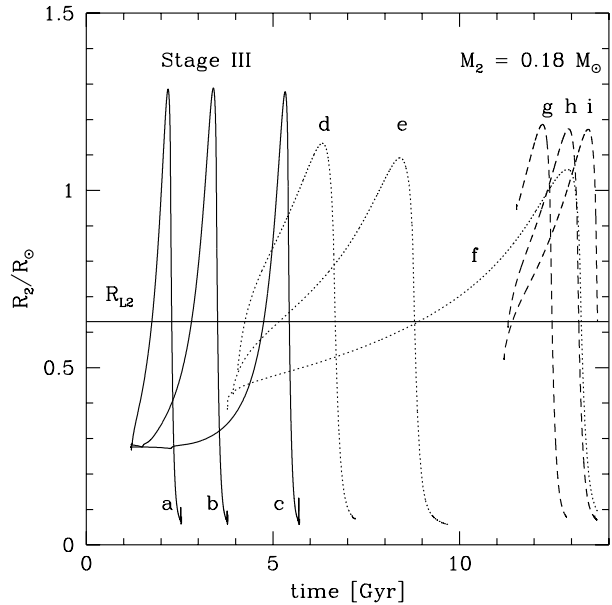}
\caption{
The time variations of the AW UMa secondary's radius
for ${\rm M_{2} = 0.18 ~ M_{\odot} }$. The different line styles 
represent its different initial masses of the secondary: solid lines  for 
${\rm M_{2,0} = 1.79 ~ M_{\odot} }$, dotted lines for 
${\rm M_{2,0} = 1.28 ~ M_{\odot} }$, and dashed lines for 
${\rm M_{2,0} = 1.00 ~ M_{\odot} }$. For each line style three tracks
are shown, corresponding (from right to left) to initial models more
and more hydrogen exhausted during Stage I (see Table 1).
 ${\rm R_{L2}}$ stands for the secondary's Roche lobe radius.}
\end{figure} 

\begin{figure}
\includegraphics[width=\linewidth]{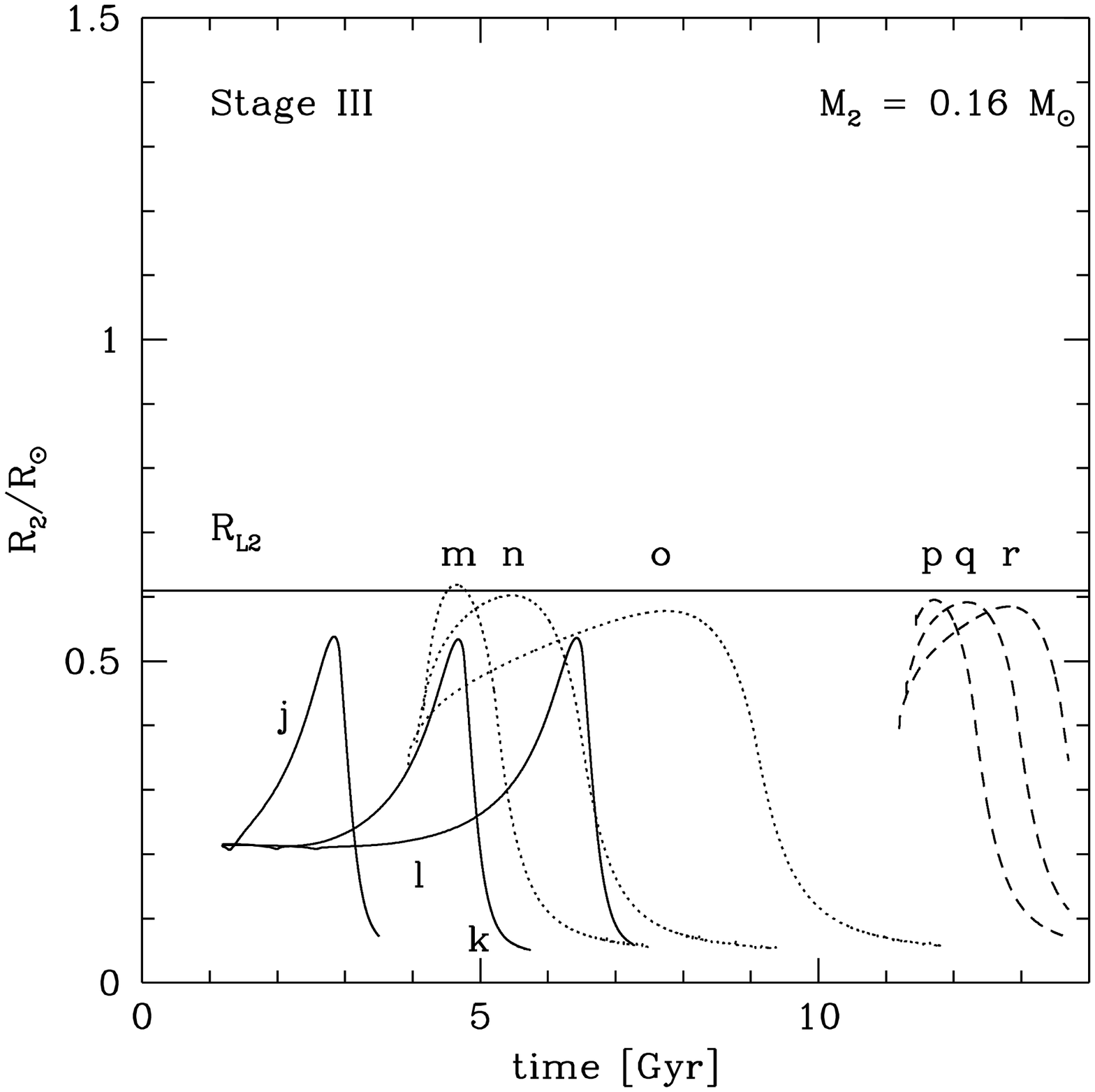}
\caption{
The same as in Fig.1 but for ${\rm M_{2} = 0.16 ~ M_{\odot} }$.}
\end{figure} 

\begin{figure}
\includegraphics[width=\linewidth]{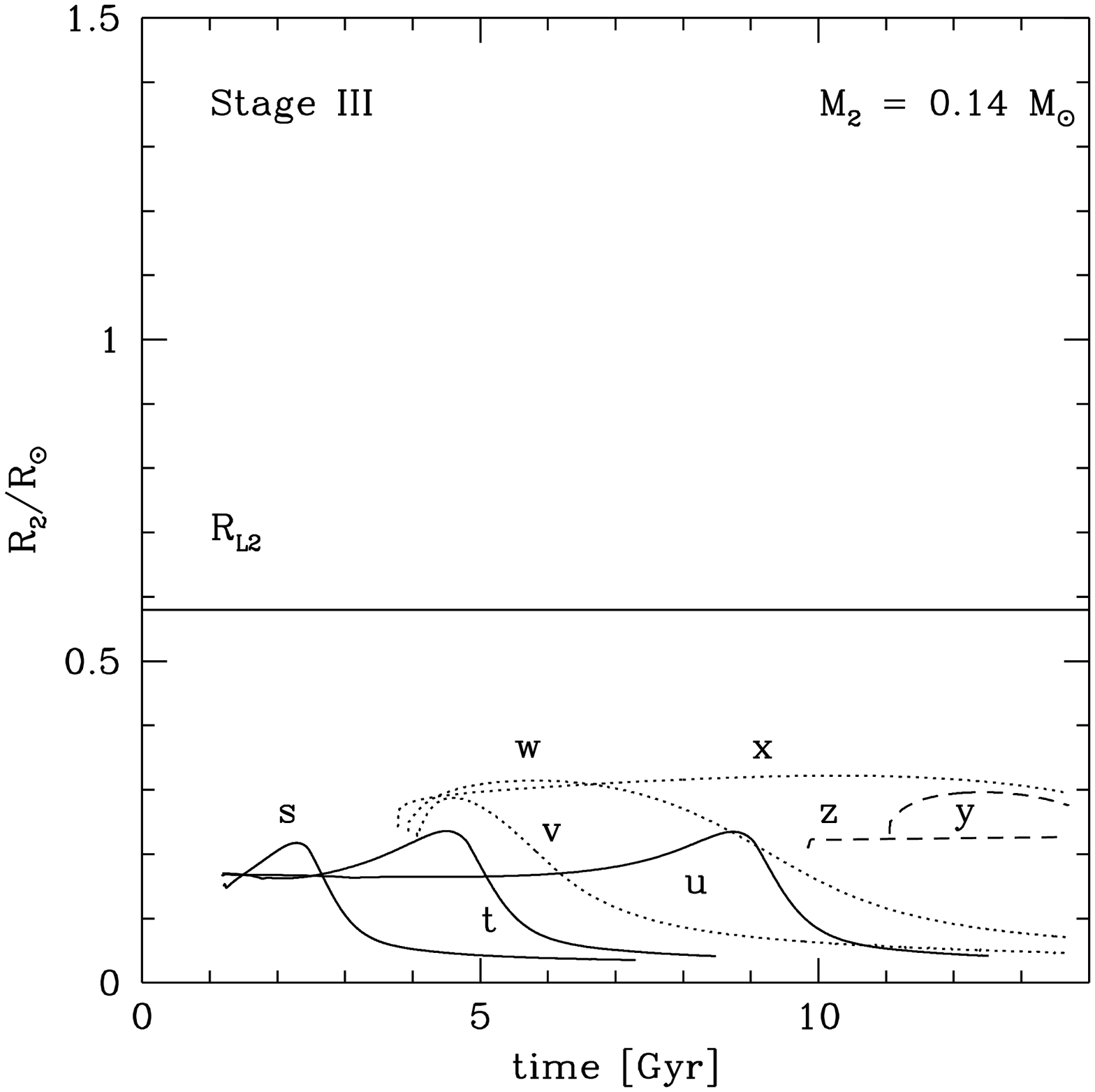}
\caption{
The same as in Fig.1 but for ${\rm M_{2} = 0.14 ~ M_{\odot} }$.}
\end{figure} 

\begin{figure}
\includegraphics[width=\linewidth]{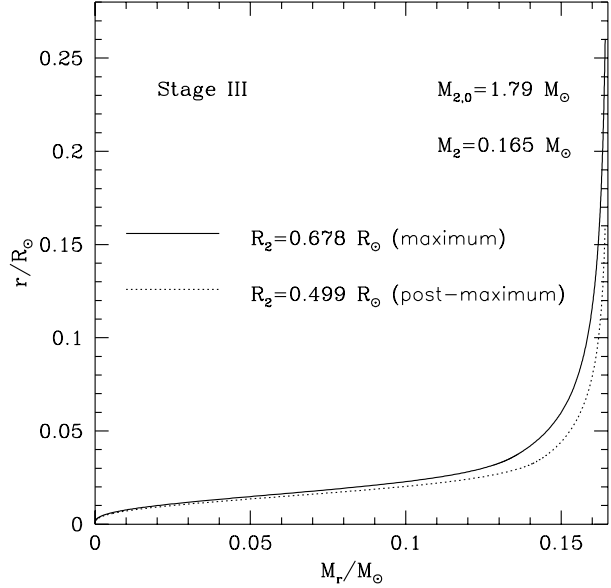}
\caption{
An example of internal structure of the AW UMa secondary.
In this figure, the internal mass-radius dependence is shown.
The track parameters are, ${\rm M_{2,0} = 1.79 ~ M_{\odot} }$,
                          ${\rm M_{2} = 0.165 ~ M_{\odot} }$,
                          ${\rm X_{c,0} = 0.014 }$. 
The model with maximal radius during Stage III
(denoted with "maximum") has been chosen to represent the internal 
structure of the low mass AW UMa component (the secondary).
 A slightly more evolved 
model ("post-maximum") shows the direction of further evolutionary changes.
For given binary period and assuming the primary mass 
${\rm M_1 = 1.61 ~ M_{\odot}}$ and 
${\rm M_2 = 0.165 ~ M_{\odot}}$ the Roche lobe 
radius is ${\rm R_{L2} = 0.613 ~ R_{\odot}}$.
}
\end{figure} 

\begin{figure}
\includegraphics[width=\linewidth]{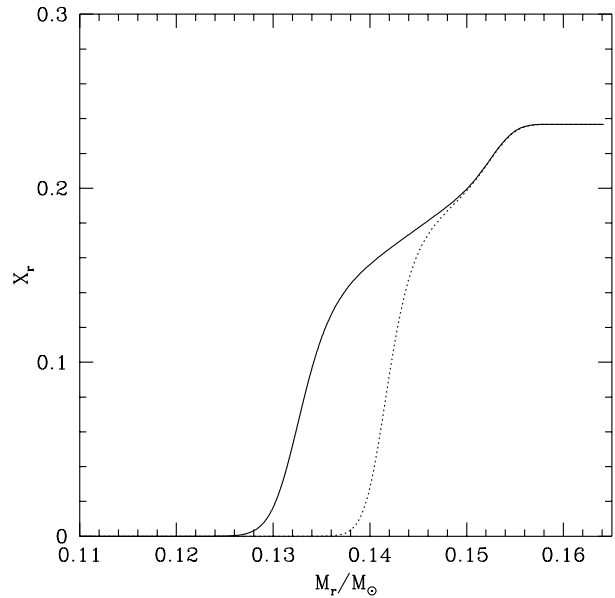}
\caption{
The same as in Fig.4 but for the internal mass-hydrogen content relation.
}
\end{figure} 

\begin{figure}
\includegraphics[width=\linewidth]{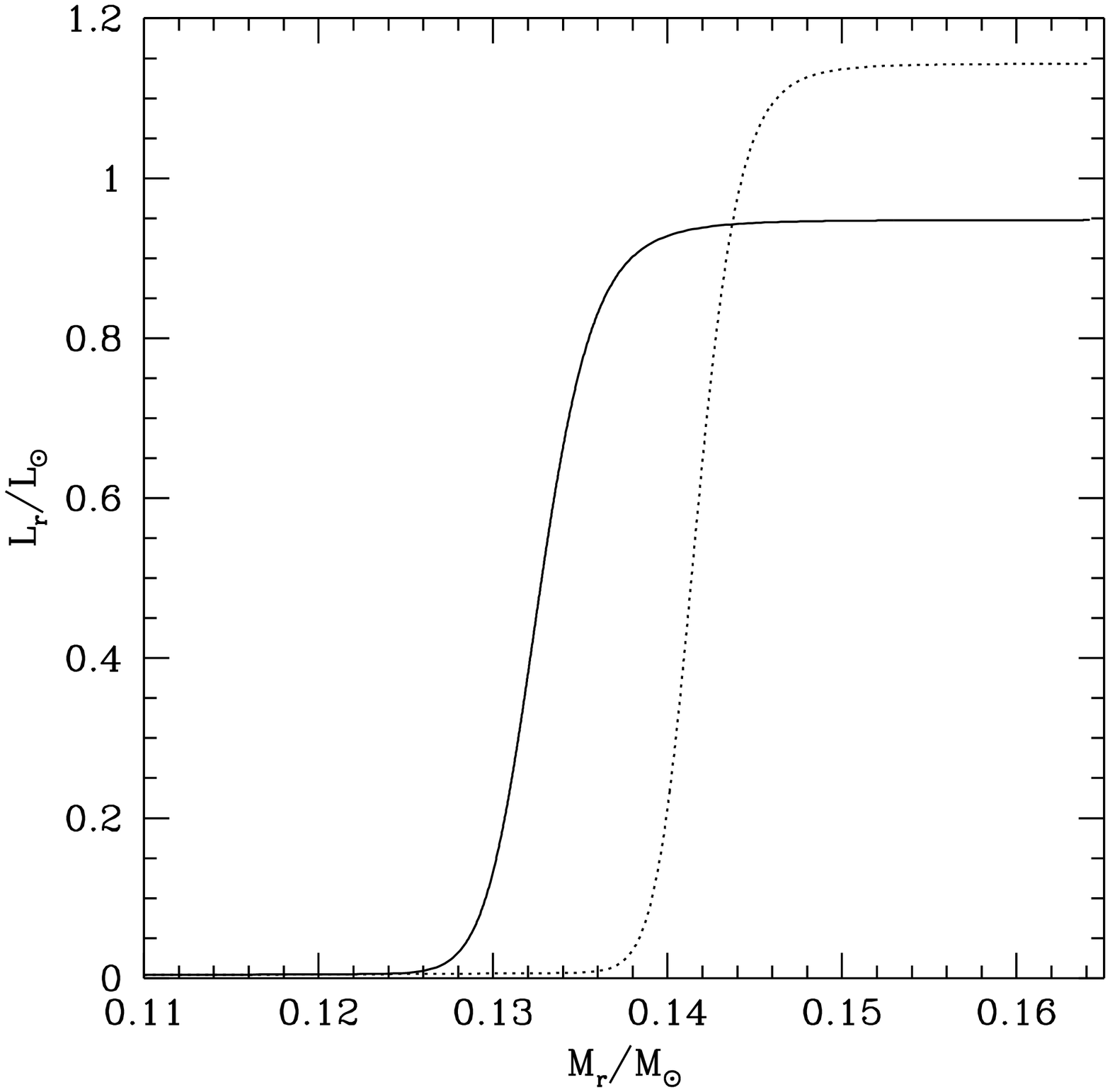}
\caption{
The same as in Fig.4 but for the internal mass-luminosity relation.
}
\end{figure} 

\section{Discussion}
\label{sect:discuss}

The most important result in this analysis are Figs. 1-3 showing three possible 
scenarios of the secondary's component evolution during Stage III,
after we have reduced its mass to the values 
${\rm M_{2}/M_{\odot} = 0.18, ~ 0.16, ~ 0.14 }$, respectively.
Depending on the initial mass of the secondary ${\rm M_{2,0}}$ and its
chemical composition at the end of Stage I, Stage III will start at various
 instants and continue for significantly different time intervals.

We see that only for ${\rm M_2=0.18 ~ M_{\odot}}$ (Fig. 1) does the secondary's
radius ${\rm R_{2}}$ exceed the secondary's Roche lobe radius ${\rm R_{L2}}$ 
(solid horizontal line) during the star's evolution, so the secondary
 is therefore able to overflow its Roche lobe and form a common envelope.
This occurs for the highest value of ${\rm M_{2} = 0.18 ~ M_{\odot}}$. 
For masses ${\rm M_{2} \leq  0.16 ~ M_{\odot}}$ the secondary will never 
overflow its Roche lobe (Fig. 2 and 3). This, combined with our estimation of the 
primary mass ${\rm M_{1} = 1.61 ~ M_{\odot}}$, requires the mass ratio for AW UMa
to be ${\rm q \geq 0.1}$.

Such a star, having a very small mass for its radius is almost an ``empty'' 
object. Its internal structure is represented in Figs. 4-6. We arbitrarily 
chose the initial mass of the secondary to be 
${\rm M_{2,0} = 1.79 ~ M_{\odot}}$ and its present mass (i.e., the secondary mass)
${\rm M_{2} = 0.165 ~ M_{\odot}}$, enough to ensure the formation of 
a common envelope. For a given binary period and assuming the primary mass 
${\rm M_{1} = 1.61 ~ M_{\odot}}$ 
the Roche lobe radius of the secondary is ${\rm R_{L2} = 0.613 ~ R_{\odot}}$.
Solid lines in these figures represent internal structure of the model
at the time when it has a maximum radius ${\rm R_{2} = 0.678 ~ R_{\odot}}$ during 
Stage III of evolution.
Dotted lines show the internal structure after some time, in a more 
evolutionary advanced stage.

When looking at the radius versus mass dependence (Fig.4) we see that matter in the 
the secondary's interior is highly concentrated towards the center resulting
in a very dense and small core and very extensive low density envelope.
The effect is that the bulk of the secondary appears as if it was almost empty.

The hydrogen inside the secondary is exhausted in its core (Fig. 5) and stays
only in an outer shell. The thickness of the hydrogen layer will decrease during 
further evolution of the secondary.
As we see in Fig.6, nuclear energy generation is limited to a very thin (in mass)
hydrogen burning shell which becomes even thinner during further evolution
of the secondary.

Our models indicate that a model of AW~UMa and similar to very 
low mass-ratio binaries involving a highly evolved secondary
component encounters a limitation at the very low ${\rm q}$ end in that
such secondaries cannot be made smaller than ${\rm 0.165 ~ M_{\odot}}$ or
thereabouts. 

Further evolution of such systems will depend mainly on the evolution of the more 
massive component. The secondary may become a degenerate dwarf and/or be 
consumed by the primary. In this scenario the system might evolve into a FK Com type 
star, as earlier proposed by Bopp and Ruci\'nski (1981), Bopp and Stencel (1981).

\section{Acknowledgments}

We are very grateful to P.P. Eggleton, D.C. Fabrycky for many helpful discussions  
and, especially, to S. Ruci\'nski and  K. St\c epie\'n for suggesting many 
improvements in a preliminary version of this paper. \\
NSF grant AST-0607070 and NASA grant NNG06GE27G are also acknowledged. 
The work of DS was supported by the MNiSW grant N203 007 31/1328.


\bsp
\label{lastpage}

\end{document}